# Stability and diffusion of oxygen vacancies in $LaNiO_3$: a DMFT study

Uthpala Herath,* Soumya S. Bhat, and Aldo H. Romero
*Department of Physics and Astronomy, West Virginia University, Morgantown, WV 26506-6315, USA*

Vijay Singh† and Hyowon Park‡
*Department of Physics, University of Illinois at Chicago*

Manipulating oxygen vacancies in strongly correlated rare-earth nickelate perovskites ($RNiO_3$) enables the tuning of their elusive metal-insulator transition (MIT), providing a better handle for control over their electronic properties. In this paper, we investigate the effect of various oxygen vacancy configurations on the MIT of $LaNiO_3$ by studying their spectral functions and the corresponding diffusion energy path using dynamical mean field theory (DMFT) and density functional theory plus U (DFT+U). To consider all possible configurations for a fixed vacancy concentration, we use a symmetry-adapted configurational ensemble method. Within this method, we can reduce the configurational space which needs to be considered, thus lowering the computational cost. We demonstrate that controlling the oxygen vacancy position can tune the occurrence of MIT. We also show that the nudged elastic band (NEB) energy barrier heights and energy profile obtained using DMFT are lower and different than those obtained using DFT+U due to dynamical quantum fluctuations among non-degenerate correlated orbitals not properly treated in DFT+U.

## I. INTRODUCTION

The metal-to-insulator transition (MIT) is an intriguing material science phenomenon that is still not quite well-understood [1, 2]. As the name suggests, in a typical MIT, there is an abrupt change in the material conductivity as it undergoes a transition from a metallic state to an insulating state. The MIT in a material can be externally stimulated through various means, including strain, doping, and vacancy creation. This transition could benefit a wide range of applications such as switching devices, Field Effect Transistors (FETs), and photonics [3]. As switches operate as states of either binary 1's or 0's, this could be manifested as either a metallic or insulating state in a material. By controlling external stimuli, this switch can be turned on or off [4]. As the transition can be made to have hysteresis, it could also act as memory storage. Hence the application in non-volatile switching devices can find use in fields such as neuromorphic computing [5]. As the transition is affected by numerous degrees of freedom, it is difficult to pinpoint the exact mechanism responsible for it. It requires a deeper theoretical understanding of the phenomena. Materials that display MIT have a critical quantum point in the electronic phase diagram representation. They could undergo a metal-to-insulator phase transition with a slight change to their atomic environment.

Though many classes of materials exhibit the MIT behavior, rare-earth nickelates ($RNiO_3$) stand out as they are more sensitive to undergoing MIT through external stimuli [6, 7]. However, the origins of this transition in these materials are elusive and still require gaps in knowledge to be filled. They can be tunable using external parameters including strain, doping, electric field, and temperature. At a critical temperature $T_{MI}$, $RNiO_3$ compounds exhibit an electronic MIT, which also involves a structural transition from $Pbnm$ to monoclinic $P2_1/n$ symmetry [8–10].

Among rare-earth nickelates, $LaNiO_3$ portrays remarkable characteristics. This material was the only known rare-earth nickelate that is a paramagnetic metal not exhibiting MIT at any temperature in bulk form [11–13]. However, recent experiments have shown that when single crystal $LaNiO_3$ are used instead of powdered or polycrystalline samples, highly conductive transport properties and magnetic ordering under high oxygen pressures are observed. Beyond the well known paramagnetic phase of $LaNiO_3$ they observed AFM configuration with an ordering temperature of $T_N$=157 K [14]. Additional studies have shown that ultrathin $LaNiO_3$ films and creating defects in $LaNiO_3$ result in MIT at room temperature [15, 16]. Another reason to study oxygen vacancies in materials lie in the fact that, unlike pure theoretical calculations, experiments performed in labs will most likely have defects such as vacancies and theoretical studies would have to account for these in order to reproduce experimental observations accurately. Therefore, it is important to gain a clear understanding of such defects in materials in order to control their effects.

Defects could be introduced through multiple means such as lattice distortion, doping or creating vacancies. Through oxygen vacancies, excess electrons are released into the environment and based on whether the material is $p$-type or $n$-type oxide it may drive a metal-to-insulator transition by reducing the hole concentration or an insulator-to-metal transition by increasing the electron concentration, respectively, at a quantum critical

* ukh0001@mix.wvu.edu
† Also at Department of Physics and Astronomy, West Virginia University, Morgantown, WV 26506-6315, USA
‡ Also at Materials Science Division, Argonne National Laboratory, Argonne, IL, 60439, USA

point. As rare-earth metals are becoming a center of attention for renewable energy applications, investigating the effect of oxygen vacancies on their electronic properties would be a timely study.

Nickelates consist of a $NiO_6^{3-}$ octahedra linked at their corners with $R^{3+}$ cations, in our case $La^{+3}$, which can be deformed through external stimuli which affect the conductivity among other properties (Fig. 1). Due to the rotations of the octahedra, the Ni-O-Ni bond lengths vary from their original 180° bond angle. The phase transitions occur primarily due to the interaction between the Ni-3d and O-2p electrons which is further elucidated from their density of state (DOS) plots which display strong hybridization between these orbitals [8]. Applications of $LaNiO_3$ include ferroelectric capacitors and non-volatile memory [17, 18]. Although there have been studies of oxygen vacancies performed using DFT and DFT+U, the application of dynamical mean field theory (DMFT) to them has been rare except for the vacancy ordered structure [19].

Although Density Functional Theory (DFT) [20, 21] is successful in treating weakly correlated materials due to the interplay between $d$ and $f$ electrons, a more accurate theory is required to capture the physics in such materials properly. This fact is due to the use the exchange-correlational term (XC) used in Kohn-Sham DFT uses a static approximation to compensate for the correlation effects arising from these localized orbitals. As an alternative, DMFT [22, 23] is a method that is now gaining popularity due to its success in studying strongly correlated materials. Effects of oxygen vacancies have been studied using DMFT previously, albeit in rare occasions, in materials such as $SrTiO_3$ and several $RNiO_3$ [19, 24, 25]. However, our study focuses on a more intricate and a systematic approach to generating and investigating the effects of vacancies. Using DMFT to probe for MIT would shed more light on its elusive behavior as it captures the dynamics of the electronic correlations which goes beyond the static limitation of DFT or DFT+U methods.

In addition to the general electronic properties study of oxygen vacancies in $LaNiO_3$, we also perform Nudged Elastic Band (NEB) calculation to investigate the energy barrier of a single oxygen vacancy diffusion as there is no such study that has been done prior to our study. Energy barrier studies are used in a variety of fields, for instance, to study the energy barrier of Li ion diffusion in battery materials [26]. By gaining sufficient knowledge on the diffusion of Li, such troubles could be avoided. In the scope of this work, oxygen vacancy in diffusion could be used in applications related to ultra-fast switching devices. In this paper, we discuss the MIT of $LaNiO_3$ studies with DFT, DFT+U and DFT+DMFT using our DMFTwDFT code [27]. Recently, Liao et al. studied $LaNiO_{2.5}$ using DFT+DMFT, and interestingly, the authors revealed that in $LaNiO_{2.5}$, the Ni octahedron site develops a Mott insulating state with strong correlations as the Ni d-$e_g$ orbital is half-filled, while the Ni square-planar site with apical oxygen vacancies becomes a band insulator.[19] However, compared to such previous studies are done to study the effects of oxygen vacancies in $LaNiO_3$ [19, 28], we use a slightly different method to generate the oxygen vacancies, which is based on Site Occupation Disorder (SOD) [29] such that the total vacancy configurational space is addressed.

This article is organized as follows. We first discuss the computational methodology followed to generate the vacancies and perform the DFT+DMFT calculations for vacancies and vacancy diffusion. Then we discuss the results focusing on stability of each vacancy case and the emergence of metal to insulator transitions. Finally, we discuss about the NEB energy barriers for both symmetry conserved and non-symmetry conserved DMFT calculations for different double counting mechanisms.

## II. METHODOLOGY

### A. Oxygen vacancy generation using the site occupation disorder (SOD) method

The number of all possible vacancy configurations in a supercell increases exponentially with the number of created vacancies. Calculating the energetics of such supercell configurations is a highly computationally expensive task, even if it could be technically possible. In order to mitigate this issue, we employed the site occupation disorder method as available through the SOD package [29]. This method reduces the number of site-occupancy configurations by utilizing the crystal symmetry of the lattice. The equivalence between configurations is gauged through *isometric transformations* which are geometric operations including translations, rotations, reflections etc. that keep all the distances and angles constant within the transformed object. Through this transformation, we obtain a *reduced configurational space* containing a significantly smaller sets of configurations compared to those with the total independent configurations.

For this study, we used the primitive 10-atom cell of $LaNiO_3$ with the $R\bar{3}c$ symmetry group in a paramagnetic structure, then generated the pristine supercell of $La_6Ni_6O_{18}$, which can accommodate different oxygen vacancies. We calculated there to be 1, 8, 28, 100, and 256 different ways of creating the oxygen vacancies for 1, 2, 3, 4 and 5 vacancies, respectively, as summarized in Table I. We realized that using the SOD method significantly reduces the configurational space. In this work we limit our electronic structure calculations to 1, 2 and 3 vacancies as the configurational space beyond that rises drastically deeming the calculations to be computationally very expensive to study through DMFT. The spacegroup symmetries of each of these supercell configurations are given in the SM. Fig. 1 and Fig. 2 of the SM show the pristine and vacancy configurations for 1 and 2 vacancies, respectively. The dotted circles represent the generated oxygen vacancy.

Table I: The number of symmetrically possible combinations for different number of vacancies as calculated with the SOD package.

| Vacancies | Supercell Combinations | Inequivalent Combinations | Formula | Reduced Formula |
|---|---|---|---|---|
| 0 | 1 | 1 | $La_6Ni_6O_{18}$ | $LaNiO_3$ |
| 1 | 18 | 1 | $La_6Ni_6O_{17}$ | $LaNiO_{2.83}$ |
| 2 | 153 | 8 | $La_6Ni_6O_{16}$ | $LaNiO_{2.66}$ |
| 3 | 816 | 28 | $La_6Ni_6O_{15}$ | $LaNiO_{2.5}$ |
| 4 | 3060 | 100 | $La_6Ni_6O_{14}$ | $LaNiO_{2.33}$ |
| 5 | 8568 | 252 | $La_6Ni_6O_{13}$ | $LaNiO_{2.16}$ |

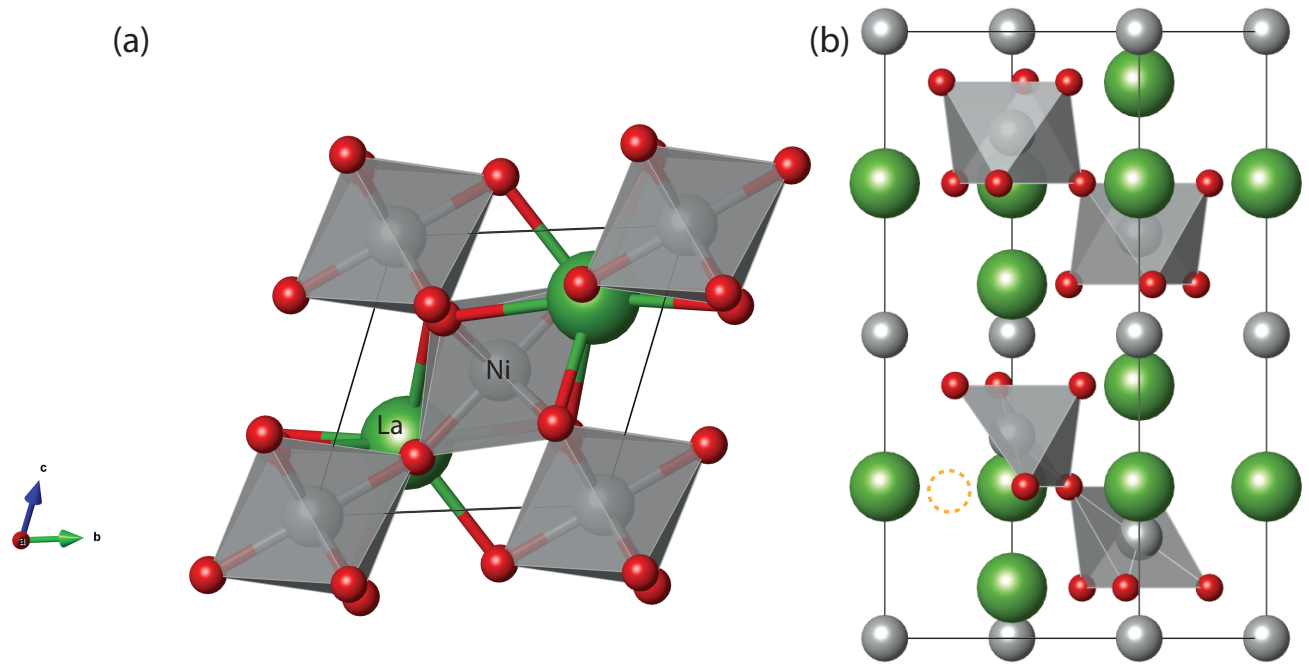

Figure 1: (a) Primitive unit cell of $LaNiO_3$ (b) A single oxygen vacancy of $LaNiO_{2.83}$ is displayed with the dotted circle. La, Ni and O are displayed in green, gray and red, respectively.

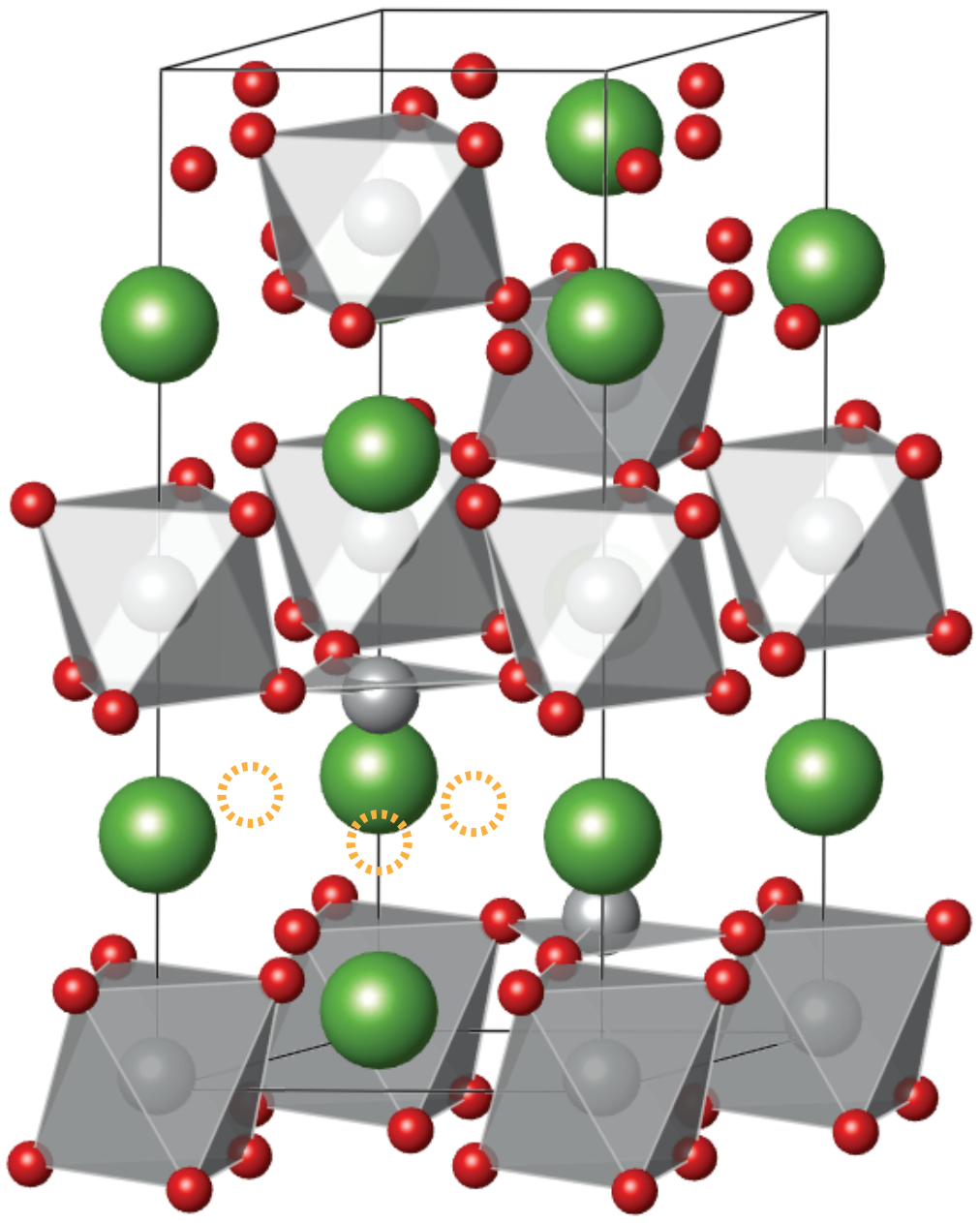

Figure 2: A single configuration of a triple vacancy structures of $LaNiO_{2.5}$. The oxygen vacancy is shown in orange dotted circles. La, Ni and O are represented in green, gray and red, respectively.

## B. DFT+DMFT calculation

We used the Vienna Ab-initio similation package (VASP) [30] to perform the DFT calculations in a combination of our in-house DMFTwDFT framework [27] to perform the DMFT calculations.

For the DFT calculation, we used the Perdue-Burke-Ernzerhof (PBE) exchange-correlation functional within the PAW method and a k-point mesh of 8×8×8 for the pristine structure and 5×5×5 for all the vacancy structures for the summation over the Gamma centered k-mesh with an energy cut-off of 692 eV. A Gaussian smearing of 0.2 eV was used. The Hubbard parameter was set to U = 5 eV and Hund's coupling was set to J=1 eV as obtained from [31] for both DFT+U and DMFT calculations. Although bulk $LaNiO_3$ is known to mostly be paramagnetic, some studies suggest four magnetic configurations namely NM, FM, AFM-I and AFM-II [32] as depicted in Fig. 3 of the SM. Therefore, we have considered the structural relaxation using each of these configurations.

Initially, the structural relaxation (both cell and atoms) was performed on the $La_6Ni_6O_{18-x}$ ($x$ = 0, 1, 2, 3) supercell configurations using DFT and DFT+U such that the force convergence criteria was satisfied at $1{\times}10^{-3}$ $eV/\AA$ per atom. The conjugate gradient algorithm was used for the relaxation.

The Kohn-Sham wave functions resulting from the DFT calculation were projected onto Maximally Localized Wannier Functions (MLWF) through the Wannier90 package [33] to construct the correlated subspace for DMFT calculations. From the DFT+U projected DOS, we noticed that there is a strong Ni-d, O-p hybridization within the energy range of -8 eV to 3.5 eV with respect to the Fermi energy level. Therefore, both Ni-d and O-p orbitals were considered in the hybridized Wanner subspace while the Ni-d orbitals are treated as correlated orbitals using DMFT. Once the Wannier function are calculated, an additional unitary transform $\hat{\Lambda}$ representing the rotations of orbitals in the correlated subspace is performed in order to minimize the off-diagonal matrix elements within each site sector of the local correlated manifold and hence minimize the off-diagonal components of $\Sigma_{loc}$. This sets the z-axis to be relative to the surrounding coordination octahedron. We used a double counting term $\alpha$ = 0.2 eV in the modified Fully localized limit (FLL) limit methods, dc_type = 1, dc_type = 2 and dc_type = 3 as implemented in the DMFTwDFT code [27]. The choice of $\alpha$ is motivated by previous work performed on $LaNiO_3$ by Park et al. [34, 35]. More explanations about different double-counting formula are given below.

In pristine $LaNiO_3$, the nominal configuration of Ni-d is $d^7$ with $t_{2g}$ fully filled and $e_g$ bands partially filled. The crystal field splitting in a $NiO_6$ octahedron separates the energy level between $e_g$ and $t_{2g}$ orbitals. This is not the case in the vacancy induced cases as the $NiO_6$ octahedron is distorted resulting in a breaking of the symmetry. In principle, all Ni-d orbitals can be non-degenerate under the low-symmetry structure. However, we performed DMFT calculations for both the Ni-d degenerate and non-degenerate cases. The Wannierization was done using a Wannier k-mesh of 24×24×24. The Wannier Hamiltonian was then mapped onto an impurity model which was solved using the hybridization expansion version of the numerically exact continuous time QMC (CTQMC) method implemented by Haule et al. [36]. We set the temperature for the calculation to be 0.01 eV. 100 one-shot DMFT iterations were run to obtain a decent self-energy convergence, i.e. when the local/lattice self-energy, $\Sigma^{cor}(i\omega)$ approaches the impurity self-energy, $\Sigma^{imp}(i\omega)$. The effective DFT+DMFT energy functional $\Gamma$ can be formulated using the four operators ($\hat{\rho}$, $\hat{V}^{Hxc}$, $\hat{G}^{cor}$, and $\hat{\Sigma}$) as shown in Eq.(2) of Ref. [27]. Here, $\hat{V}$ is the double-counting potential operator along with $E^{DC}$, the double counting energy.

Once the self-consistency conditions are reached, the DFT+DMFT total energy is calculated as Eq. (1) from the Free energy functional, $\Gamma$ in the zero temperature limit.

$$E = E^{DFT}[\rho]+\frac{1}{N_{\mathbf{k}}}\sum_{i\mathbf{k}}\epsilon_i^{\mathbf{k}}\cdot\left(n_{ii}^{\mathbf{k}}-f_i^{\mathbf{k}}\right)+E^{POT}-E^{DC} \quad (1)$$

Here, $E^{DFT}[\rho]$ is the DFT energy calculated with the DFT charge density, $\epsilon_i^k$ is the DFT KS eigen value, $n_{ii}^k$ is the diagonal of the DMFT occupancy matrix $n^k$, and $f_i^k$ is the Fermi function (DFT occupancy matrix) with the KS band $i$ and the momentum $\mathbf{k}$.

The potential energy $E^{POT}$ is evaluated through the Luttinger-Ward functional, $\Phi[G^{cor}]$ is given by the Migdal-Galistkii formula [37].

$$E^{pot} = \frac{1}{2}\,\mathrm{Tr}\left[\Sigma\cdot G^{cor}\right] = \frac{1}{2}\sum_{\omega_n}\left[\Sigma\left(i\omega_n\right)\cdot G^{cor}\left(i\omega_n\right)\right] \quad (2)$$

Here, $E^{pot}$ is calculated using the Migdal-Galisky formula and one can also obtain it using the direct sampling of CTQMC. We find that both methods produce very similar results.

The double-counting energy term in Eq. (1) can be evaluated through multiple means which are implemented in the DMFTwDFT framework. The fully localized limit (FLL) is used widely in DFT+U calculations and is given by,

$$E^{DC} = \frac{U}{2}\cdot N_d\cdot(N_d-1)-\frac{J}{4}\cdot N_d\cdot(N_d-2) \quad (3)$$

Previous studies have shown that the conventional double-counting energy is overestimated for the metal-insulator boundary, d-p spectra and energetic compared to experiments [38–44]. Therefore, we have introduced a fine tuning parameter $\alpha$ to improve this through multiple dc types.

- dc_type=1:

$$E^{DC} = \frac{(U-\alpha)}{2}\cdot N_d\cdot(N_d-1)-\frac{J}{4}\cdot N_d\cdot(N_d-2) \quad (4)$$

- dc_type=2:

$$E^{DC} = \frac{U}{2}\cdot(N_d-\alpha)\cdot(N_d-\alpha-1)-\frac{J}{4}\cdot(N_d-\alpha)\cdot(N_d-\alpha-2) \quad (5)$$

In Eq. (4), $E^{DC}$ is reduced by the lowering of $U$ by $\alpha$ term and in Eq. (5) $E^{DC}$ is reduced by the lowering of $N_d$ by $\alpha$. Setting $\alpha = 0$ recovers the original FLL term in Eq. (3). Additionally, we have implemented dc_type=3 which uses the nominal occupancy of the correlated orbital, $N_d^0$ [43] and is independent of the $\alpha = 0$ parameter. The double-counting potential for this case is given by,

$$V^{DC} = \frac{U}{2}\cdot N_d^0\cdot\left(N_d^0-1\right)-\frac{J}{4}\cdot N_d^0\cdot\left(N_d^0-2\right) \quad (6)$$

For DMFT post-processing, the self-energies were analytically continued onto the real axis using the maximum entropy method [45] which is essentially $\Sigma(i\omega)\to\Sigma(\omega)$. This is necessary since CTQMC samples the self-energy on the imaginary axis. This self-energy was then used to perform further post-processing including calculating the spectral function $A(\omega)$ and DMFT density of states. The spectral function is given in Eq. (7).

$$A(\omega) = -\frac{1}{\pi}\ln G(\omega) \quad (7)$$

Here, $G(\omega)$ is the local Green's function. To obtain the DMFT band structure the Eq. (8) was used.

$$A(k,\omega) = -\frac{1}{\pi}\frac{\mathrm{Im}\,\Sigma(\omega)}{\left(\omega-\epsilon_k-\mathrm{Re}\,\Sigma(\omega)\right)^2+\left(\mathrm{Im}\,\Sigma(\omega)\right)^2} \quad (8)$$

Additionally, we calculated the mass renormalization ($m^*/m_0$) by calculating the inverse of quasi-particle residue $Z$ given by,

$$Z^{-1} = 1-\left(\frac{\partial\,\mathrm{Re}\,\Sigma(\omega)}{\partial\omega}\right)_{\omega=0} = \frac{m^*}{m_0} \quad (9)$$

### C. Nudged Elastic Band calculations for single oxygen vacancy diffusion

The NEB method is a popular method for calculating the minimum energy pathways of kinetic processes. In this work we have used NEB through VTST tools [46–48] along with DiSPy [49] to study the energetics involved in oxygen vacancy diffusion for a single oxygen vacancy. A wrapper developed by Romero Group, NEBgen (https:

//github.com/uthpalaherath/NEBgen) was used to automate the complete process to generate symmetrically unique paths for the single oxygen vacancy system through Distortion Symmetry Method with DiSPy and then run the NEB with VTST Tools. Although, linear interpolation between an initial and final structure (image) provides a decent estimate for the minimum energy pathway, the Distortion Symmetry Method takes into account symmetry-adapted perturbations to systematically lower the initial path symmetry, enabling the exploration of other low-energy pathways that may exist. VTST uses VTST scripts to perform a linear interpolation between the first and last image and performs the NEB calculation with VASP once the Distortion Symmetry Method is applied to the images with DiSPy. However, with the preliminary calculations we performed, we realized that due to the lower symmetry (P1) of the vacancy structure, the distortion paths generated by DiSPy and VTST were identical. Therefore, we only performed NEB calculations using the linear interpolated path using VTST tools. The following setup for NEB calculations were used. A NEB climbing image algorithm with 7 intermediate images and a spring constant of 5.0 was used. The Force optimizer used was *quick-min* (IOPT=3) with a EDIFFG=-0.05 and 1000 NSW steps. We obtained a NEB force convergence to the order of 1 eV/Å. The NEB was performed with DFT+U. The resulting images (structures) of the NEB path were then used to perform DMFT calculations to obtain the DMFT total energy which was used to generate the DMFT energy profile for the vacancy diffusion.

## III. RESULTS AND DISCUSSION

### A. Structural relaxation

Following the introduction of oxygen vacancies, we performed a structural relaxation as discussed in Sec. II. The relaxations were performed with NM, FM, AFM-I and AFM-II magnetic ordering imposed with an initial magnetic moment of 2 $\mu_B$ on each Ni atom [32]. We also noticed that using a higher initial magnetic moment such as 5 $\mu_B$ resulted in the same ground states.

The total energies of these ground states are shown in Table II. For the single vacancy structure, the AFM-II and FM ordering was found to be the ground state of DFT and DFT+U, respectively. For the double vacancy, the AFM-I state was found to be the lowest energy while with DFT+U it was found to be in FM, similar to the DFT+U ground state of the single vacancy. For the triple vacancy, it was found that the AFM-II phase was the ground state with both DFT and DFT+U. For double and triple vacancies, the total energies were averaged over their number of possible configurations i.e. 8 and 28, respectively. These outcomes are consistent with several experimental and computational studies [19, 50, 51]. Although the magnetic moments after the relation was much lower than their initial value of 2 $\mu_B$, we did not see any consistent pattern of these magnetic moments based on the proximity of their atom to the vacancy site. In the presence of oxygen defects, the possible geometries for the Ni-O bonding environment are shown in Fig. 3. For the pristine structure, all the Ni atoms form a $NiO_6$ octahedra with equal Ni-O bond lengths. Based on the total energies, it was evident that the most stable cases for the oxygen vacant structures were ones that had the most number of square-planar Ni-O geometries (Fig. 3(d)). In fact, for the double vacancy case, out of the 8 possible configurations, only one had a square-planar geometry and this configuration had the lowest total energy. For the triple vacancy case, there were 28 possible configurations and the three with the highest number of square-planar geometries (two square-planar (Fig. 3(d)) each, along with two octahedra (Fig. 3(a)) and two half-octahedra (Fig. 3(b)), had the lowest three total energies. The next stable configurations consisted of at most one square-planar geometry with a mixture of octahedra and half-octahedra environments. With this, we could anticipate that the higher number of square-planar geometries in oxygen vacant structures lead to higher stability. This outcome is in consistent with studies done by Refs. [19, 28]. Additionally, configurations containing more quarter-octahedra Fig. 3(c) and trigonal-planar Fig. 3(e) type geometries have higher ground state energies.

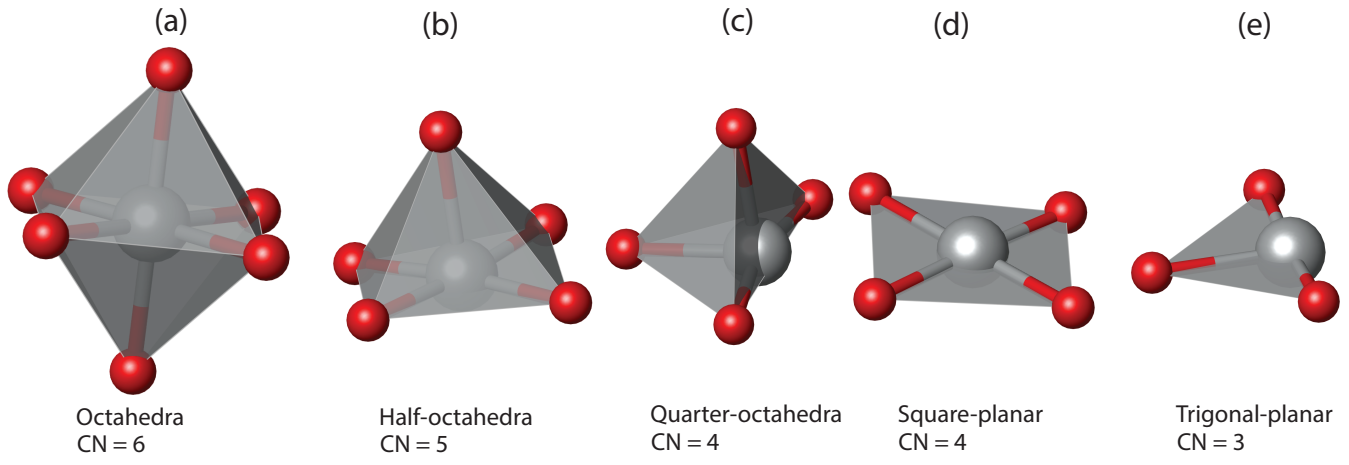

Figure 3: The possible Ni-O coordination geometries in $LaNiO_3$. The Ni and oxygen ion are displayed in grey and red, respectively. The coordination numbers for each geometry are given by **CN**.

Table II: The total energy of different magnetic configurations for single, double and triple oxygen vacancies with respect to their ground state. The total energy for double and triple vacancies are averaged over their number of different configurations. The initial magnetic moment used for the relaxation was 5 $\mu$B. The U and J values for DFT+U are 5 eV and 1 eV, respectively.

| Ordering | $\Delta E_{TOT}$ (eV) | | | | | |
|---|---|---|---|---|---|---|
| | 1vac | | 2vac | | 3vac | |
| | DFT | DFT+U | DFT | DFT+U | DFT | DFT+U |
| NM | 0.0130 | - | 0.1970 | - | 0.1507 | - |
| FM | 0.0134 | 0 | 0.1940 | 0 | 0.8552 | 0.2254 |
| AFM-I | 0.0173 | 0.2310 | 0 | 0.1990 | 0.7698 | 0.1485 |
| AFM-II | 0 | 0.4170 | 0.7350 | 1.0630 | 0 | 0 |

For the DMFT calculations following the structural relaxation, we used the optimized NM structures in each case assuming a paramagnetic (PM) configuration to include the dynamical correlation effect. As a final note, we record that the energy difference between the lowest energy state and the highest energy state were $\sim$ 2.20 eV, $\sim$ 2.88 eV and $\sim$ 25.03 eV per supercell, calculated with DFT, DFT+U and DMFT, respectively. A comparison of these DFT, DFT+U and DMFT energy distributions and the metallic/ insulating nature of each of the triple vacancy configurations is given in Table II of the SM.

### B. Goldschmidt tolerance factor

The Goldschmidt tolerance factor (t), of a perovskite is given by Eq. (10), and is a measure of the stability and distortion of a perovskite [52].

$$t = \frac{r_A + r_O}{\sqrt{2}\,(r_B + r_O)}, \tag{10}$$

where $r_A$, $r_B$, $r_O$ is the radius of the A cation, B cation and oxygen atom, respectively. If t=1 it means the atoms fit perfectly in the given structure. If t>1 it means that the A cation is too large and if t<0.8 it means it is too small to fit in. In both these cases, this mis-fitting causes distortion in the structure in the form of octahedral rotations. Bulk $LaNiO_3$ has a tolerance factor, t=0.97 [53] signalling a decent stability. The Ni-O bond increases when $Ni^{+3}$ goes towards a $Ni^{+2}$ oxidation state decreasing the tolerance factor. The Shannon's radii, which measures ionic radii for different coordination numbers (CN), for $Ni^{+3}$ and $Ni^{+2}$ are shown in Table III [54]. Additionally, the larger Ni-O bond narrows the bandwith of conduction and valence bands. This creates a stronger coulomb repulsion and possibly can lead to a large Mott-Hubbard splitting.

Table III: The Shannon ionic radii of $Ni^{+2}$ and $Ni^{+3}$ ions.

| Ion | CN | Ionic radius (nm) |
|---|---|---|
| $Ni^{+2}$ | 4 | 0.55 |
| | 4-SP | 0.49 |
| | 5 | 0.63 |
| | 6 | 0.69 |
| $Ni^{+3}$ | 6* | 0.60 |

*High-spin state Ref: [54]

Our calculations for the tolerance factor for different vacancy cases is summarized in Table IV. The ground state magnetic configurations were considered for each vacancy case as calculated in Table II. As in some structures, the octahedra environment is modified, we have included the tolerance factor only for cases where a octahedra is preserved as this is required for its definition. For better accuracy, the values here are averaged over multiple bond lengths of La-O and Ni-O. Our calculated values for the pristine $LaNiO_3$ Ni-O bond length, 1.95 Å and 1.96 Å with DFT and DFT+U, respectively, are consistent with the experimental value of 1.933 Å obtained from neutron diffraction experiments [51]. We also noticed that for the double vacancy case, DFT+U does not result in any configurations with a planar geometry, however, DFT does. The tolerance factor calculated to be 0.93 for both DFT and DFT+U for the pristine $LaNiO_3$ is also similar to the value found by [53]. For the triple vacancy case which corresponds to $LaNiO_{2.5}$ the calculated Ni-O bond length for the full octahedra environment, 2.11 Å, 2.16 Å and the Ni-O bond length for the planar environment, 1.83 Å, 1.84 Å is consistent with their corresponding experimental values obtained by Ref. [50] which are 2.12 Å and 1.91 Å, respectively. According to Table IV, we notice that DFT predicts a rather stable tolerance factor with increased number of vacancies, except for the triple vacancy case where it becomes 0.87. On the contrary, tolerance factors calculated by DFT+U show a lowering of values upto the tripe vacancy case where the tolerance factor becomes 0.79. This discrepancy between DFT and DFT+U values could possibly be attributed to effects arising from correlations coupled with environment changes due to the creation of vacancies.

Table IV: Calculated value of crystal tolerance factors and bond length between La-O and Ni-O for each vacancy case with DFT and DFT+U is shown. The tolerance factors are only defined for full octahedra environments.

| Structure | DFT | | | DFT+U | | |
|---|---|---|---|---|---|---|
| | $d_{La-O}$ (Å) | $d_{Ni-O}$ (Å) | t | $d_{La-O}$ (Å) | $d_{Ni-O}$ (Å) | t |
| Pristine | 2.58 | 1.95[a] | 0.93 | 2.58 | 1.96[a] | 0.93 |
| 1vac | 2.53 | 1.97[a] | 0.91 | 2.47 | 1.98[a] | 0.88 |
| | | 1.94[b] | - | | 1.96[b] | - |
| 2vac | 2.57 | 1.97[a] | 0.92 | 2.45 | 2.02[a] | 0.86 |
| | | 1.97[b] | - | | 1.98[b] | - |
| | | 1.92[c] | - | | - | - |
| 3vac | 2.60 | 2.11[a] | 0.87 | 2.41 | 2.16[a] | 0.79 |
| | | 1.83[c] | - | | 1.84[c] | - |

[a] octahedra
[b] half-octahedra
[c] plane

### C. DMFT density of states

As expected the pristine $LaNiO_3$ was found to be metallic as seen in its spectral function and DMFT DOS depicted by Fig. 4. Since the Ni-O octahedral symmetry is conserved in this case, we preserved the Ni-$e_g$ and Ni-$t_{2g}$ degeneracy in the DMFT calculation.

However, due to the breaking of symmetry of the $NiO_6$ octahedra as oxygen vacancies are introduced, the cubic symmetry within Ni-d $e_g$ and $t_{2g}$ manifolds is broken. Therefore, we also performed the calculations using full broken symmetry, i.e. breaking both the Ni atomic and orbital degeneracy. Fig. 5 shows the DMFT DOS of the single oxygen vacancy case with broken degeneracy. There is only one symmetrically inequivalent configuration for this case. The Wannier functions for the single vacancy case compared to the pristine is illustrated in Fig. 1 in the SM. This system is metallic which is the expected behaviour as the oxygen deficiency in $LaNiO_{2.83}$ is not sufficient to drive a metal to insulator transition. Based on work done by Ref. [16, 19] the transition occurs at $LaNiO_{2.5}$. Ni-3 and Ni-6 atoms consist of a half-octahedra environment due to the creation of the vacancy while the rest retain their full-octrahedra environments, albeit with slightly different Ni-O bond lengths in contrast to the pristine case.

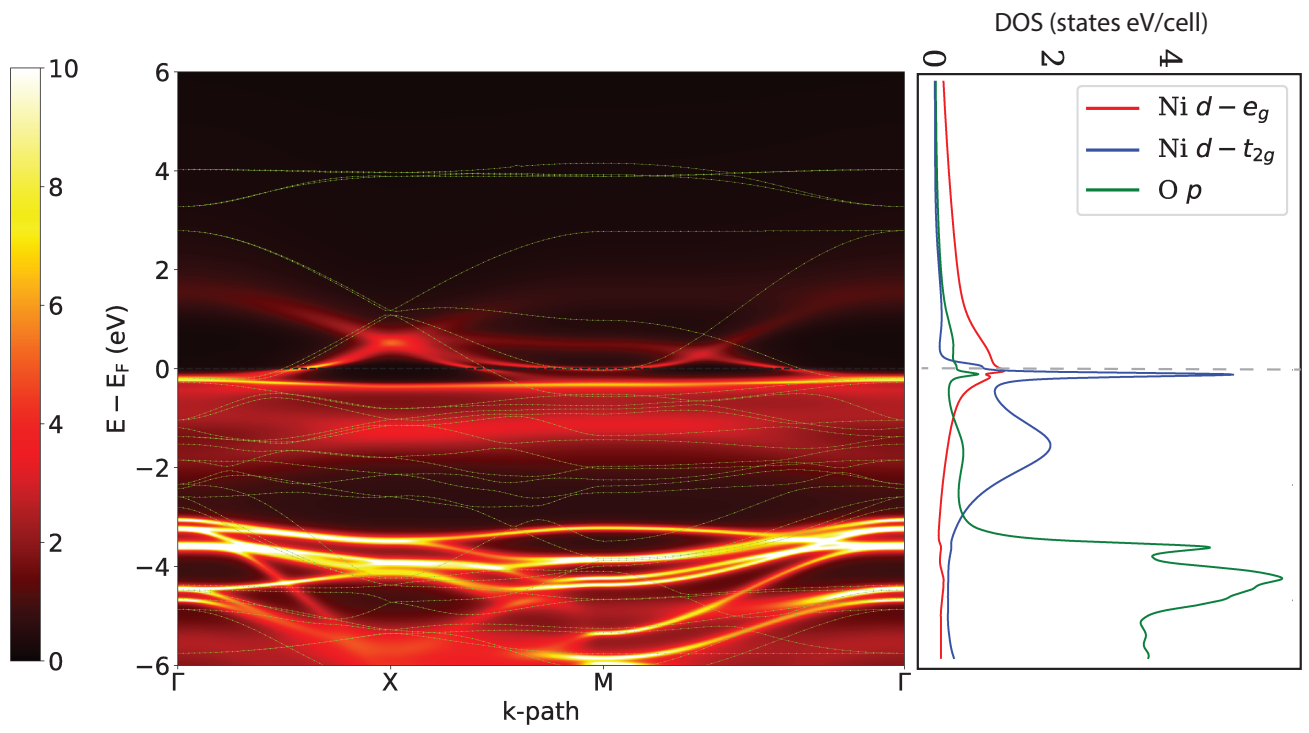

Figure 4: DFT bands (green) and DMFT bands (colormap) of pristine $LaNiO_3$ (left). DMFT DOS of pristine $LaNiO_3$ (right).

The breaking of the octahedra environment due to the introduction of an oxygen vacancy has multiple effects on the DOS. Although Ni-1, Ni-2, Ni-4, Ni-5 have their octahedra geometry intact, their Ni-O bond lengths have changed slightly which appears to be the reason why Fig. 5 shows slight degeneracy breaking in their DOS. However, in Ni-3 and Ni-6 where the octahedra environment is completely broken and resembles the half-octahedra geometry, the degeneracy breaking is much more evident. Additionally, we see that for the Ni-3 atom, all orbitals shift their position above the Fermi level, a trend also followed by Ni-6 atom, except for its Ni $d_{xy}$ orbital. Moreover, we noticed that in the octahedra preserved cases, Ni d-$e_g$ shows conducting behavior as these orbitals cross the Fermi level, whereas in the octahedra broken environments the DOS shows insulating behavior. Therefore, we may be able to use the stability of the octahedra environment as an indicator of the metallic/ insulating behavior of the DOS of those atomic environments.

The density of states for all of the 8 configurations of the double vacancy case ($LaNiO_{2.66}$) was found to be metallic for U=5 eV. This also is in agreement with previous work [16, 19]. Increasing the U value may induce a MIT due to the increase of Coulomb repulsion, however since the U value reported for $LaNiO_3$ was 5 eV, we only investigated the effects keeping it constant.

However, the DMFT DOS for the triple vacancy case is a bit more interesting as the metallic/insulating behavior varies with the configuration. Our DFT calculations showed all the configurations to be metallic which is possibly due to the failure in the XC functionals to account for correlation effects properly. Using DFT+DMFT, however, 57.1% of the configurations showed metallic behavior while the rest were insulating. The lowest energy configuration for both DFT+U and DMFT was insulating, agreeing with previous studies. For the metallic configurations we were interested in calculating the effective mass enhancement (m*/m). The reason this is calculated only for the metallic case is because quasi-particle residue is not defined for insulators. The calculated values are shown in Fig. 4 of the SM. The DMFT calculations were performed with broken d orbital and Ni atom degeneracy. The x-axis corresponds to the configuration and the y-axis corresponds to the effective mass enhancement. (a), (b), (c), (d) and (e) correspond to octahedra, half-octahedra, quarter-octahedra, square-planar and trigonal-planar, respectively, as shown in Fig. 3. In some configurations not all geometries were found, which is the reason for the configuration axis to be different.

From Fig. 4 of the SM we see that the octahedra environment is the most prevalent among the metallic configurations, while the trigonal-planar had only 4 metallic configurations. We also noticed that the Ni d-$x^2-y^2$ showed high effective mass enhancements in most of these metallic configurations closely followed by Ni d-$z^2$.

To explore along this avenue further, we also plot the imaginary part of the self-energy, Im $\Sigma(i\omega)$ along with the corresponding DOS for an insulating configuration from of the triple vacancy structures.

Fig. 6 (top) shows a wide band-gap DOS and Fig. 6 (bottom) shows the Im $\Sigma(i\omega)$ for the Ni orbitals. At $\omega = 0$ traces of strong correlations, the poles and diverging nature at $\omega = 0$, leading to a Mott insulating behavior are seen in the $e_g$ orbitals of Ni-octahedra, $e_g$, $xz$, $yz$ orbitals in the Ni-half octahedra and all the orbitals in the Ni-quarter octahedra environments. In the former, $t_{2g}$ orbitals behave as a band insulator while in the second only the $xy$ orbital retains this band insulating nature.

## D. Effect of correlations on single oxygen vacancy diffusion

In this section we discuss the Nudged Elastic Bands (NEB) [46, 48] calculations performed to calculate the energy barrier of a single oxygen vacancy diffusion in $LaNiO_3$. Although, vacancy diffusion could have been studied for a higher number of vacancies, for this study we only limited it to a single vacancy case as increasing the number of vacancies renders a very complex diffusion problem. We use DFT+U for the initial step of the NEB calculation. As the lowest energy for a single oxygen vacancy takes the FM magnetic configuration as we showed in Table II for DFT+U we use that here for both the initial and final states of the NEB path.

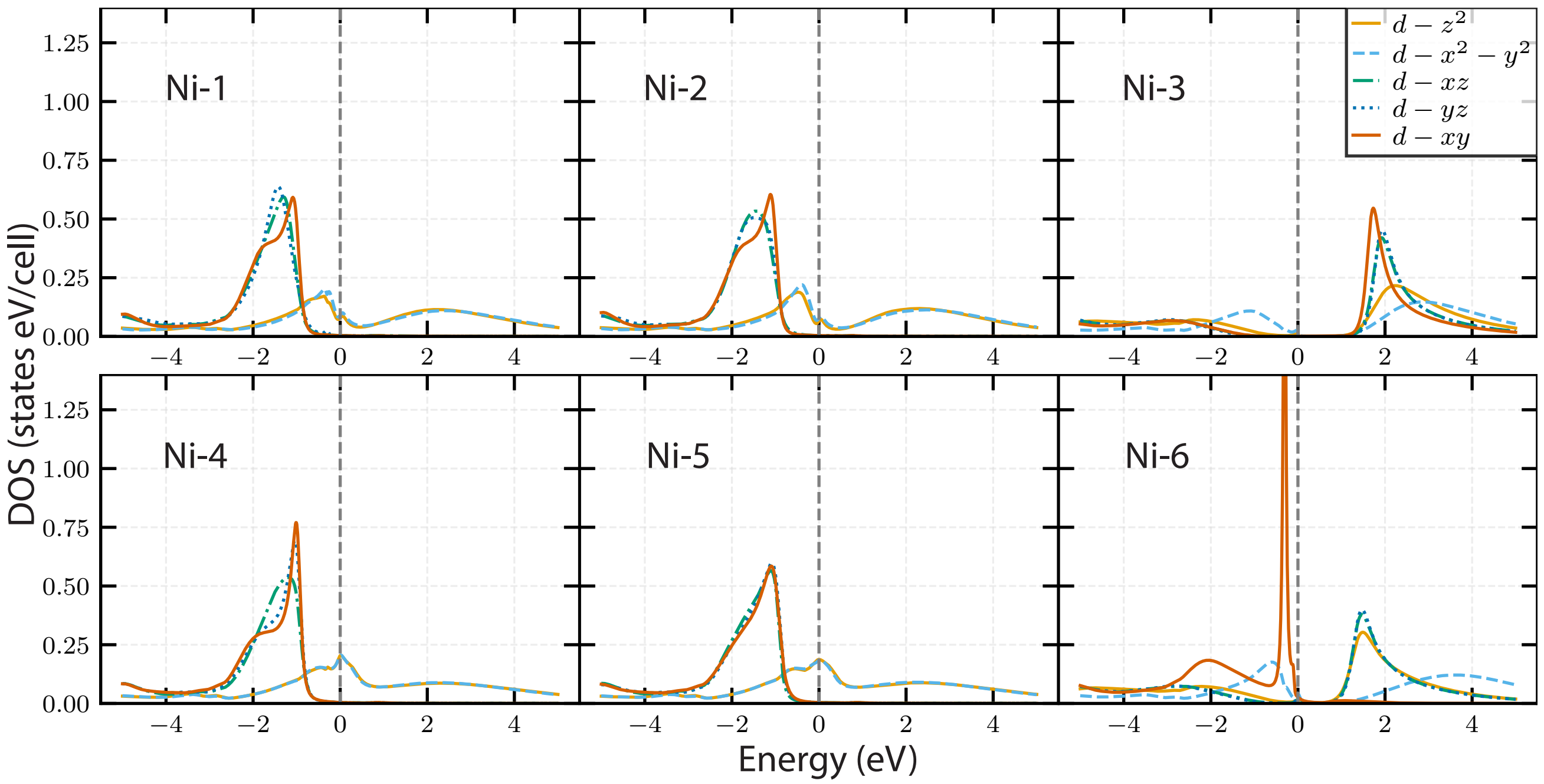

Figure 5: The DMFT PDOS of the symmetry broken Ni-d orbitals for a single oxygen vacancy. The octahedra environment of Ni-1, Ni-2, Ni-4 and Ni-5 are preserved while it is broke in Ni-3 and Ni-6. The latter shows insulating behavior while the former shows metallic given the Ni d-$e_g$ orbitals cross the Fermi level.

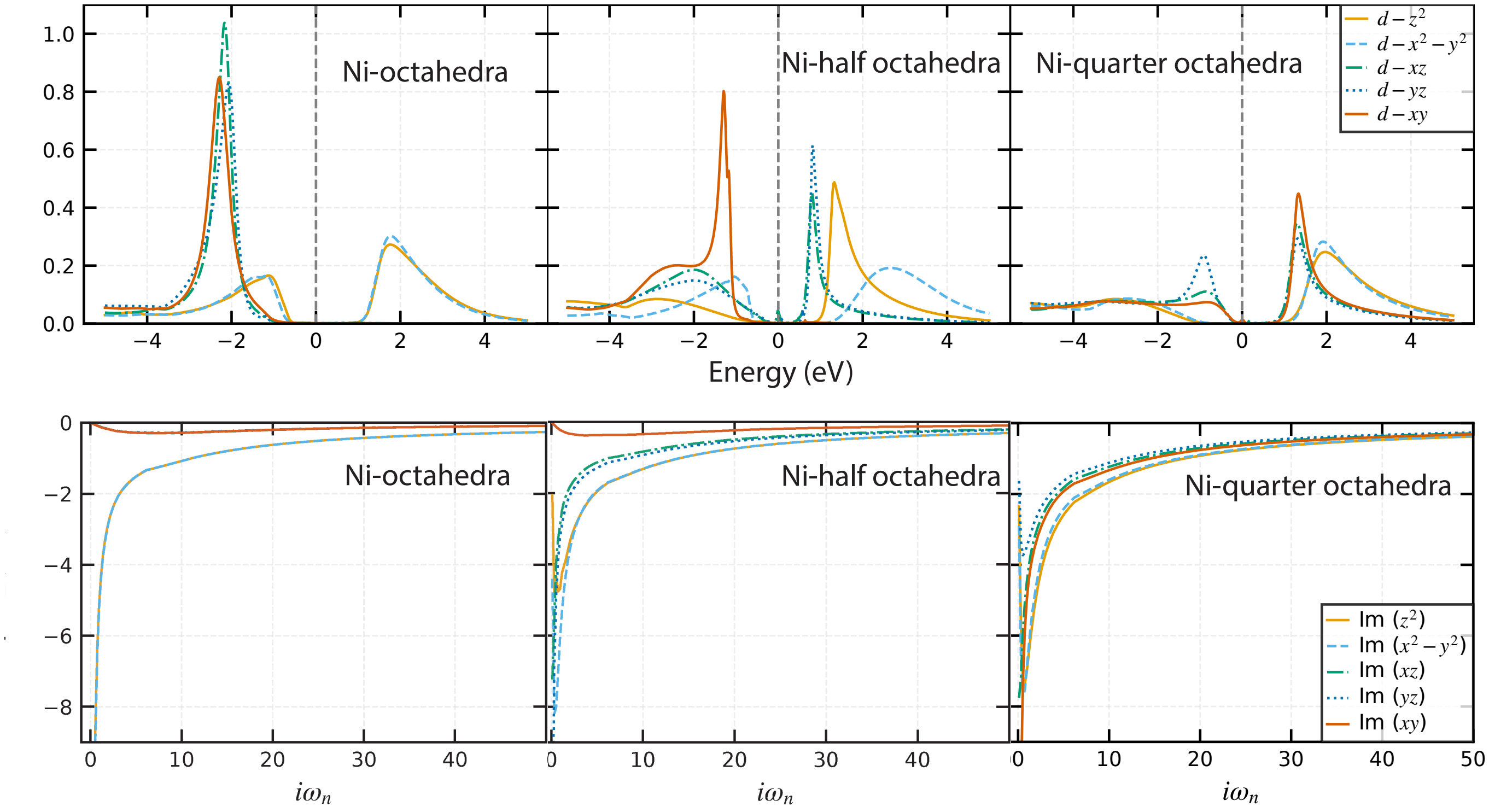

Figure 6: The DMFT DOS (top) and imaginary part of self-energy (bottom) for an insulating configuration for the triple vacancy structure. The plots are done for each available Ni-O manifold geometry for the configuration.

Our initial methodology intended to use DiSPy [49] to generate a path for the NEB calculation based on irreducible distortion symmetry, however, we noticed that the linear interpolated path between the end-points created by VTST Tools [47] is identical to the path generated by DiSPy due to the lack of symmetry in the oxygen vacant structure.

Next we performed the Nudged Elastic Band calculation with VTST Tools [47] along the generated path as outlined in Sec. II. Once the NEB energy profile using DFT+U was obtained, we used the intermediate structures along the path to perform the subsequent DFT+DMFT calculation. The parameters used for the DFT+DMFT calculation was the same as in the previous section. The DMFT calculation here was done by keeping the Ni d orbital symmetry preserved. The resulting energy profile consisting of both DFT+U and DFT+DMFT energies is shown in Fig. 7 along with key intermediate images as insets. The calculated energy barrier with DFT+U was found to be $\sim$ 0.75 eV per atom. The DMFT energy barrier was found to be 0.6 eV per atom and 0.58 eV per atom, calculated from the Migdal-Galistkii and CTQMC sampling methods, respectively using `dc_type 1`. Per atom here refers to all La, Ni and O atoms in the supercell. Malashevich et al. performed a NEB calculation for single oxygen vacancy diffusion using LDA for a 79-atom supercell and found the energy barrier height to be 1.24 eV [28]. In our study, the higher energy of the barrier height obtained with DFT+U is possibly due to the lack of properly addressing quantum fluctuations of electrons which are captured by DMFT unlike static mean field theories. However, if we used DMFT forces for the NEB instead of performing DMFT calculations with the images retrieved from the DFT+U NEB, this energy profile may have significant differences.

We used two different methods to calculate the DMFT total energy, namely, Migdal-Galistkii [37] and CTQMC sampling, which seems to have similar energy profiles. The magenta circle in the inset represents the migration of the oxygen vacancy throughout the energy profile.

However, one limitation of this case is the fact that we have performed the DMFT calculations imposing the $e_g$ and $t_{2g}$ degeneracy. To explore if this is the case we extended our DFT+DMFT calculations along the NEB path with broken degeneracy, that is to say all the d orbitals have broken symmetry for each of the 6 Ni atoms, which are also considered to be degenerate. The resulting PDOS for each of these Ni atoms are shown in Fig. 8 for the image #5 in the energy profile. This particular image was selected as it depicts the largest variation between DFT+U and DMFT profiles.

We note that the DOS of all orbitals are conducting as expected for the single vacancy case. However, we are more interested in the collective conductive properties which dictate the macroscopic properties. Therefore, we sum the contributions of each orbital for all the Ni atoms and plot the total DOS for each of the images as shown in Fig. 9 using both DFT+U and DMFT NEB results.

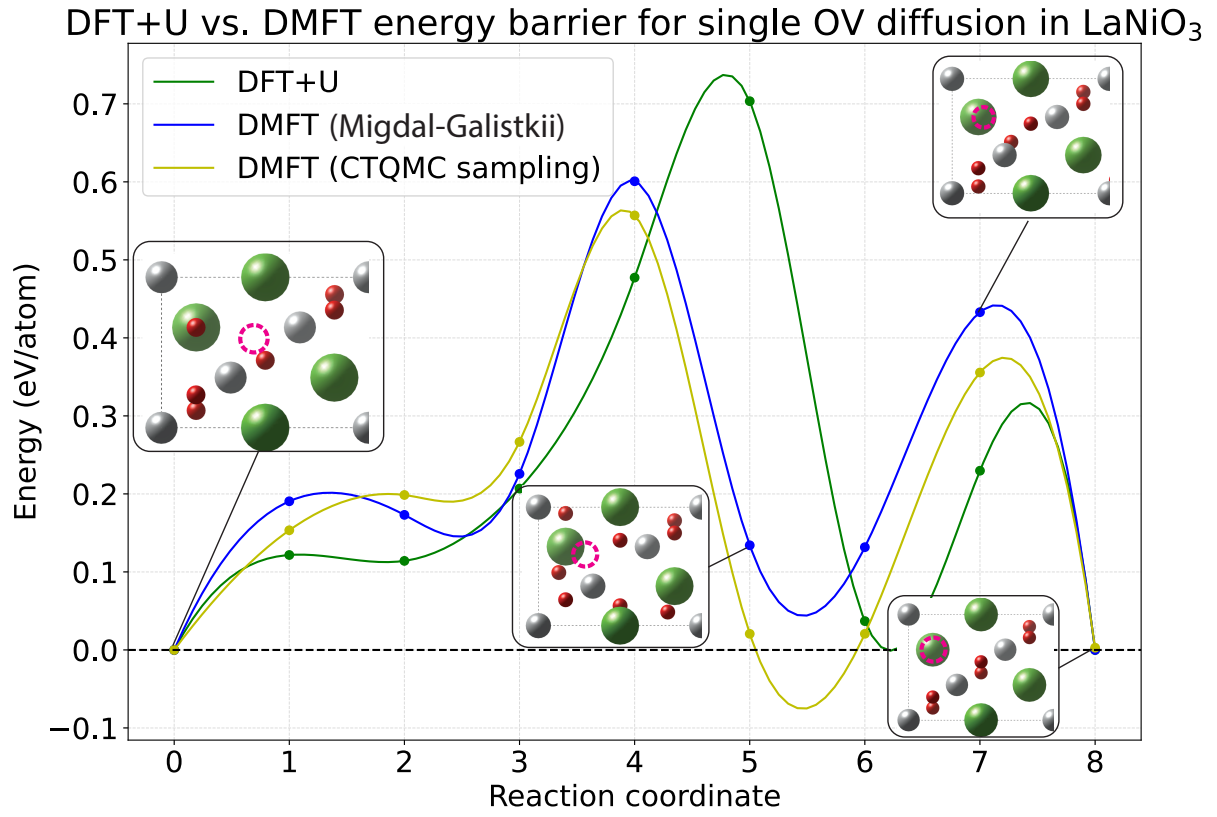

Figure 7: Minimized NEB for a single oxygen vacancy diffusion in $LaNiO_3$ calculated with DFT+U and degenerate DMFT using `dc_type 1`.

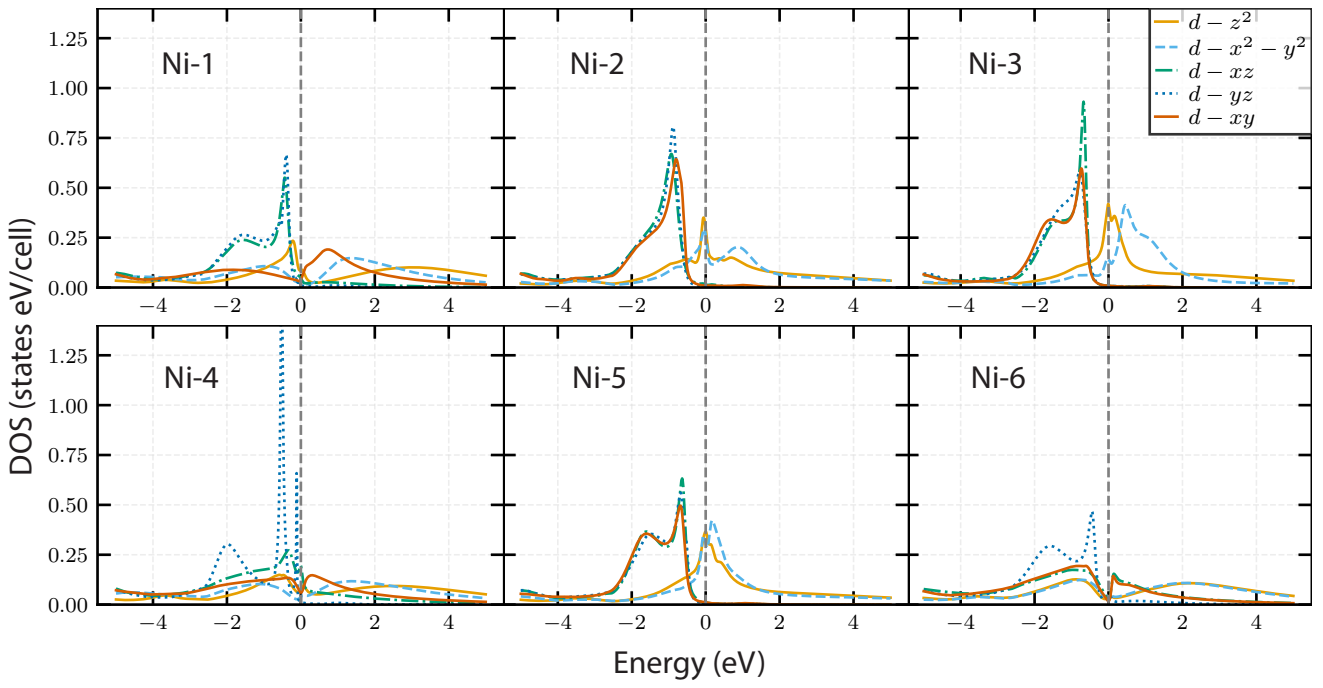

Figure 8: The DMFT PDOS of the $LaNiO_3$ NEB by breaking the degeneracy of d orbitals for each Ni atom. The DOS here are shown for image #5. `dc_type 1` was used for the DMFT calculation.

We note that the conductivity appears to differ for each image along the NEB path also depending on DFT+U and DMFT DOS variations. We notice that all the DFT+U DOS are lower than the DMFT DOS except for image #3, which DMFT predicts to be more insulating. Additionally, image #5 shows the largest variation between the DFT+U and DMFT profiles which we also noticed in Fig. 7. This makes it evident that the system changes conductivity along the vacancy diffusion path.

The outcomes of DMFT calculations depend significantly on the double counting method used. For the NEB DMFT calculations thus far in this section, we have been using the `dc_type 1` as available through the DMFTwDFT framework and calculated by Eq. (4). Next we calculated the DMFT NEB for the non-degenerate case. A comparison between the energy profiles calculated with and without using degeneracy is given in Fig. 5 of the SM. Subsequently, we consider only the CTQMC sampled DMFT energies and the non-degenerate case to compare the effects of using different double counting methods to the energy profile. The resulting energy profile comparison is shown in Fig. 10.

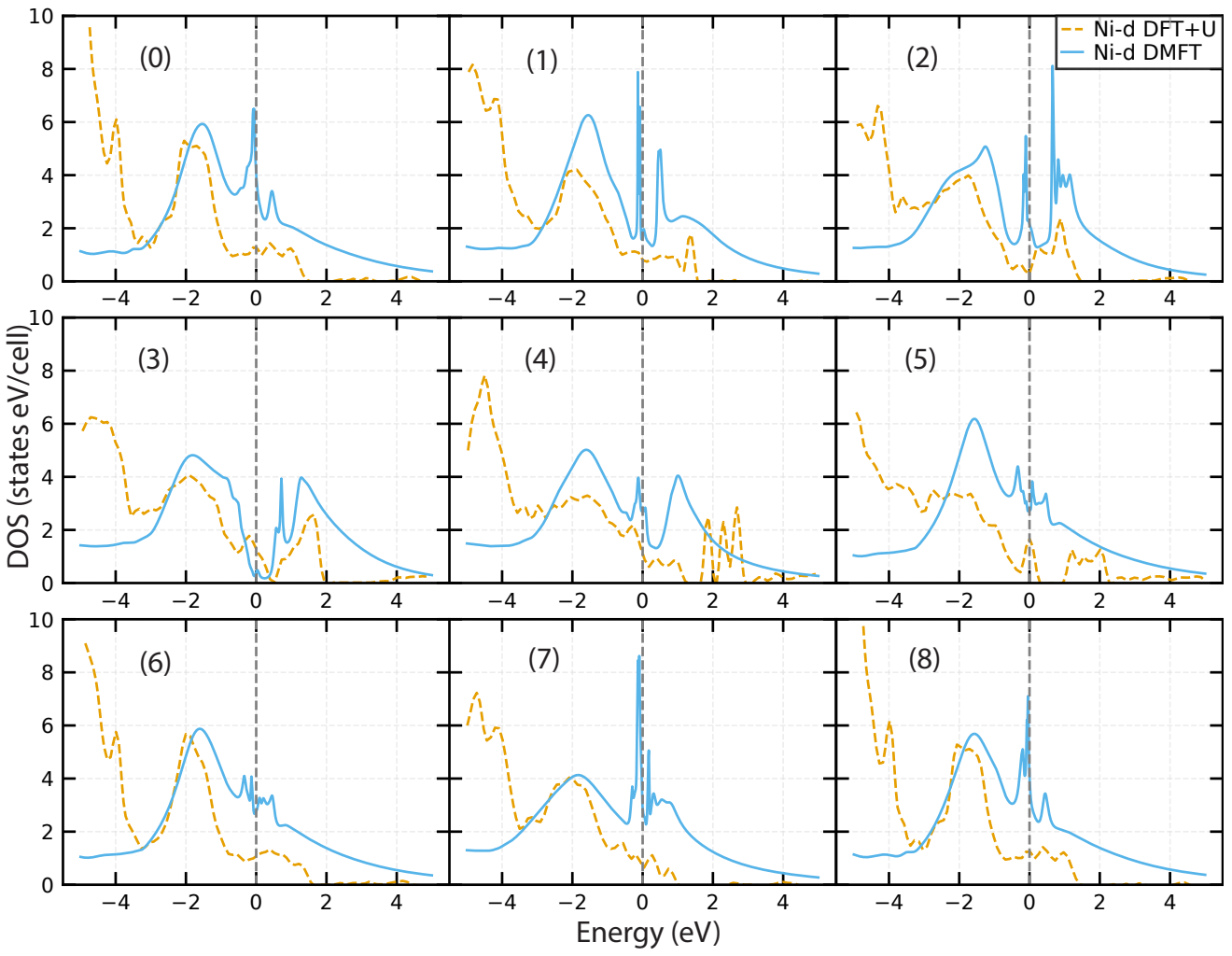

Figure 9: The DMFT total DOS of all d orbitals of each Ni atoms for non-degenreate DMFT calculation for the NEB path. `dc_type 1` was used for the DMFT calculation.

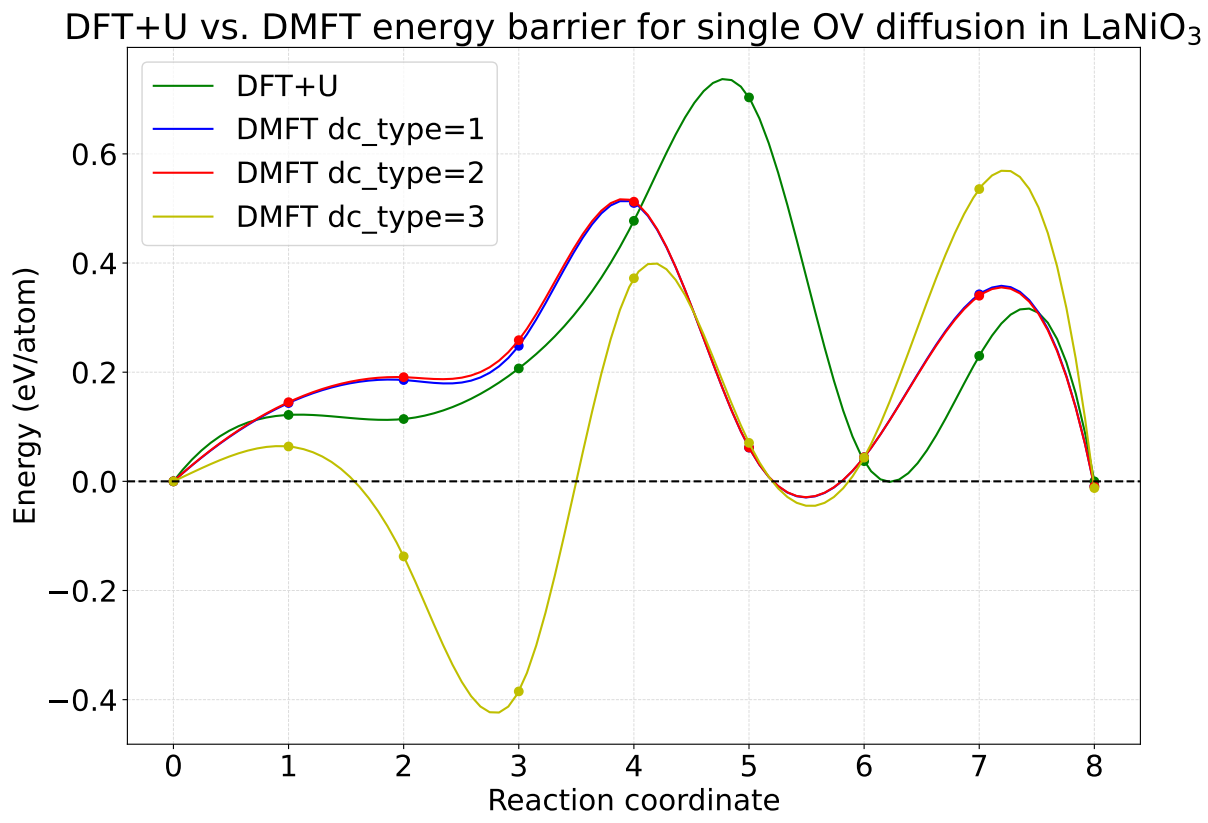

Figure 10: The NEB based on double counting type used for the DMFT calculation. The energies are obtained from non-degenerate DMFT calculations and only considering the ctqmc sampling energies. The `dc_type 1`, `dc_type 2` and `dc_type 3` are represented in blue, red and yellow curves, respectively.

There seems to be drastic differences between the preserving degeneracy (Fig. 7) and breaking it (Fig. 10). In general the non-degenerate curve for `dc_type 3` calculated with ctqmc sampling is lower in energy. It is lower in energy than the degenerate DMFT curve for the whole profile, but with respect to DFT+U, it is only lower until image #6. Between the initial state and the 4th image, the non-degenerate DMFT curve has negative energies suggesting the diffusion in this region to occur spontaneously. The reason for the non-degenerate DMFT curve to be lower in energy might be due to the fact that when the symmetry is broken it gives rise to higher degrees of freedom lowering the energy. There appears to be two peaks along this path, the first being around image #4 with a height of $\sim$ 0.8 eV and the second being around image #7 with a height of $\sim$ 1.0 eV with respect to the lowest point of the profile. These values are lower than that of the degenerate DMFT value, 1.4 eV.

The energy for `dc_type 1` and `dc_type 2` are calculated within DMFT using Eq. (4) and Eq. (5), respectively. Fig. 10 shows that `dc_type 1` and `dc_type 2` are almost identical and has a higher energy than that of `dc_type 3` along the NEB path at any point. `dc_type 1` and `dc_type 2` energy profiles do not result in negative energies as much as `dc_type 3` does. This suggests that the initial and final states of the path calculated with `dc_type 3` is unlikely.

Therefore, we can discard `dc_type 3` for this calculation and only focus on `dc_type 1` and `dc_type 2`. Comparing with the DFT+U profile, the profile of `dc_type 1` and `dc_type 2` has its highest peak around image #4 of around $\sim$ 0.52 eV and a secondary peak at image #7 of $\sim$ 0.35 eV, whereas the DFT+U profile has its primary peak which is the highest around image #5 of 0.74 eV and the lower secondary peak between images #7 and #8 of 0.3 eV which is lower than the DMFT values. A summary of the comparison of barrier height's of the primary peaks of DFT+U and DMFT energy profiles with Ni-d orbital degeneracy breaking is presented in Table V.

Further, we decomposed the DMFT total energy to potential energy and kinetic energy terms as shown in Fig. 11. The equations used here are formulated in Ref. [27]. In the images where potential energy dominates, the electrons are more localized. The DMFT kinetic energy (KE) and DMFT potential energy (PE) displayed here are independent of the double counting method used. Therefore, this plot could be used to isolate the effects of the double counting type. The total energy is calculated as a function of KE and PE along with an added term for double counting correction and from the figure we can see how this changes the energy profile with the dc type used. This further establishes that effects of `dc_type 1` and `dc_type 2` are almost identical, but `dc_type 3` is significantly different. For the barrier height changes with respect to double counting type, please refer to the SM.

Additionally, the variation of KE and PE along the diffusion path is a signature of changes to the itinerant and localization behavior throughout the diffusion process.

We also calculate the average Ni d occupancies along the energy profile for `dc_type 1` and `dc_type 3`. Our results are shown in Fig. 12. The averaging is done over all the non-degenerate Ni-d orbitals of all 6 Ni atoms. Although we did not see a direct correlation with the Ni d occupancies with the energetics of the profile, we did notice that `dc_type 3` shows lower occupancy than that of `dc_type 1`. The variation of the occupancies is due to the changes in the Ni oxidation state due to the migration

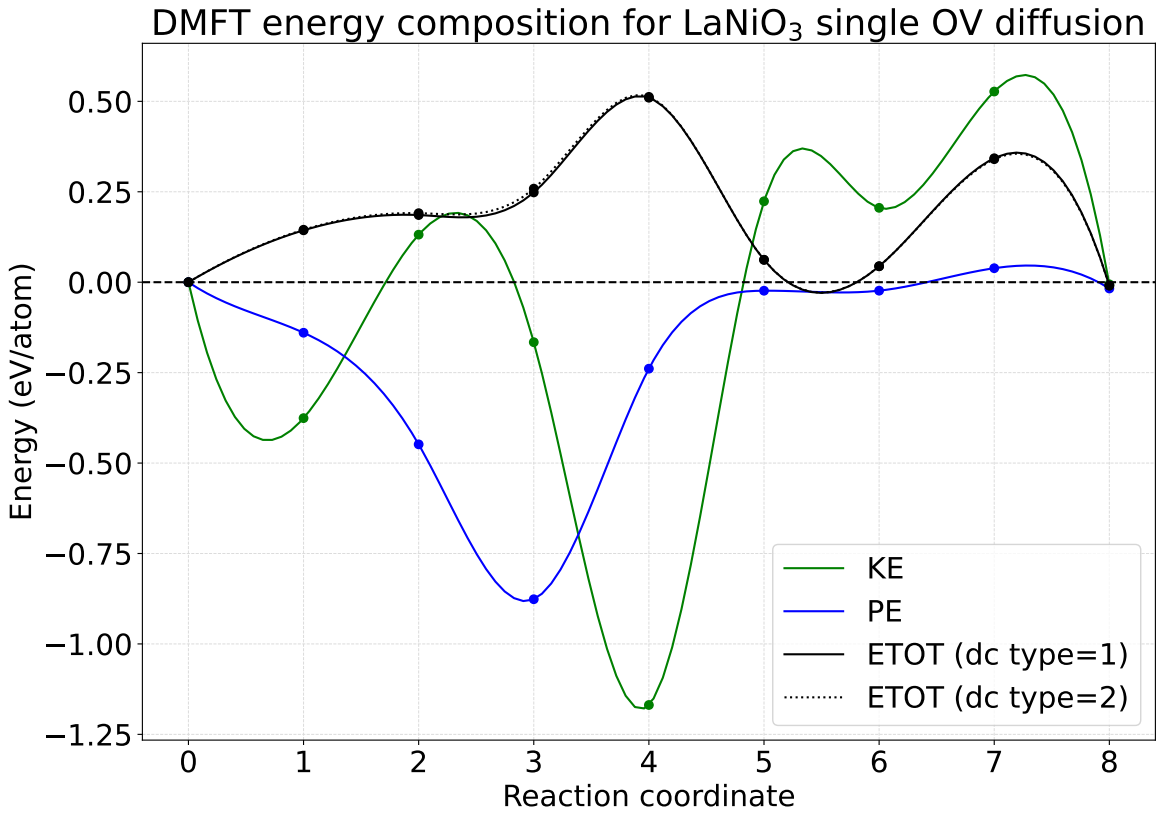

Figure 11: The total DMFT energy decomposition for the NEB path. `dc_type 1` and `dc_type 2` are displayed here in solid and dotted lines, respectively.

Table V: The barrier height's of the primary peaks of DFT+U and DMFT energy profiles for broken Ni orbital degeneracy. The DMFT total energy is computing using Migdal-Galistkii and CTQMC sampling methods. Note, we use the `dc_type 1` and `dc_type 2`.

| DFT+U | Barrier Height (eV) | |
|---|---|---|
| | DMFT (MG) | DMFT (CTQMC) |
| 0.75 | 0.40 | 0.52 |

of the oxygen vacancy along the diffusion path.

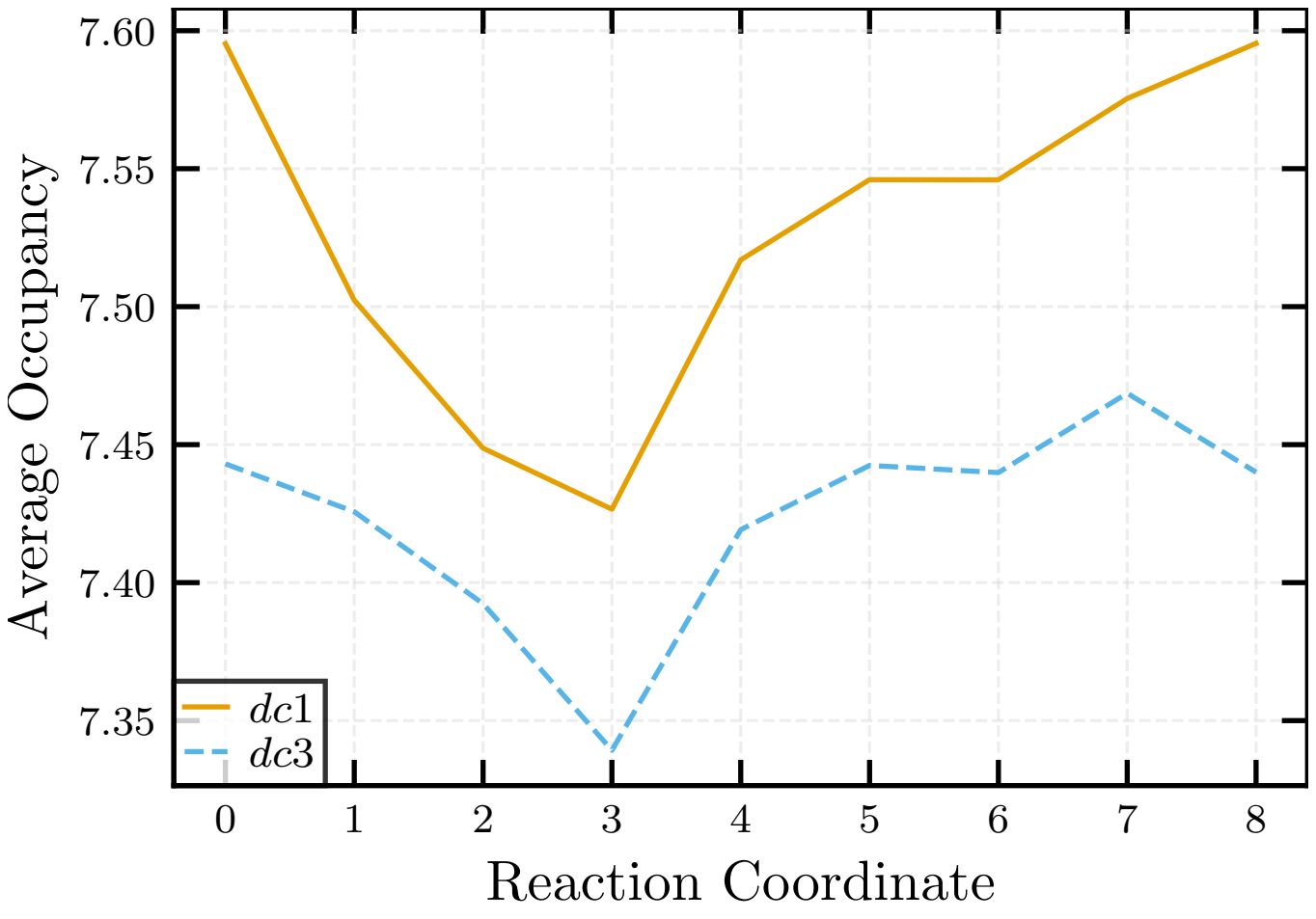

Figure 12: The variation of Ni d occupancies along the energy profile according to the `dc_type` used.

## IV. CONCLUSIONS

Introducing oxygen vacancies in $LaNiO_3$ transforms the system from metal to insulator at sufficient U and based on the vacancy configuration. We noted that for single and double vacancy, all configurations resulted in metallic behavior which is consistent with previous studies. However, for triple vacancy case, we noticed that the metallic/insulating behavior varies depending on the configuration. Based on the coordination geometry of the vacancy structures, the stability of the system can be gauged. This is an extension to the work done by Liao et al. where a single structure of the triple vacancy case, $LaNiO_{2.5}$ was studied. Additionally, we studied the effective mass enhancements of the different octahedral types for the triple vacancy case.

We studied the single oxygen vacancy diffusion energy path using NEB with the DFT+U force calculations. We also performed the total energy calculations in DMFT using the same diffusion path. Remarkably, the energy barrier heights obtained using DMFT is lower than DFT+U due to dynamical quantum fluctuations among non-degenerate correlated orbitals treated in DMFT. When this fluctuation effect is reduced by imposing the $e_g$ and $t_{2g}$ degeneracy (which is typically absent in the low-symmetry disordered structure), the energy barrier becomes higher than the non-degenerate case.

We demonstrated the effect of double counting types to the energy profile and realized the non-self-consistent nominal double counting (`dc_type 3`) did not show realistic outcomes regarding its energy profile. We also noticed that along the vacancy diffusion path, the conductivity of the system changes in addition to the localized/itinerant behaviour displayed from the KE/PE profiles. This could be a potential application in switching devices. Although, vacancy diffusion could have been studied for a higher number of vacancies, for this study we only limited it to a single vacancy case as increasing the number of vacancies renders a very complex diffusion problem.

## ACKNOWLEDGMENTS

The work was supported by the grant DE-SC0021375 funded by the U.S. Department of Energy, Office of Science. We also acknowledge the computational resources awarded by XSEDE, a project supported by National Science Foundation grant number ACI-1053575. The authors also acknowledge the support from the Texas Advances Computer Center (with the Stampede2 and Bridges supercomputers). U.H. acknowledges the WVU HPC computing facilities Spruce Knob and Thorny Flat. H. Park acknowledges funding from the US Department of Energy, Office of Science, Basic Energy Sciences, Division of Materials Sciences and Engineering.